\documentclass{aa}
\usepackage{epsfig}
\usepackage{graphicx}

\usepackage{epstopdf}
\newcommand{\eqref}[1]{(\ref{#1})}

\begin{document}

\title{Dissipative instability in partially ionised prominence plasmas}
\author{ Istvan Ballai\inst{1}, Ramon Oliver\inst{2} ,  \and Marios Alexandrou\inst{1}}
\institute{Solar Physics and Space Plasma Research Centre (SP$^2$RC), Department of
Applied Mathematics, The University of Shef{}field, Shef{}field,
UK, S3 7RH, email: {\tt i.ballai@sheffield.ac.uk}
\and Departament de F\'isica, Universitat de les Illes Balears, 07122 Palma de Mallorca, Spain, email:{\tt ramon.oliver@uib.es} 
}
\date{Received dd mmm yyyy / Accepted dd mmm yyyy}

\authorrunning{Ballai et al.}
\titlerunning{Dissipative instability in partially ionised prominence plasmas}

\abstract{}
{We investigate the nature of dissipative instability at the boundary (seen here as tangential discontinuity) between the viscous corona and the partially ionised prominence plasma in the incompressible limit. The importance of the partial ionisation is investigated in terms of the ionisation fraction. }
{Matching the solutions for the transversal component of the velocity and total pressure at the interface between the prominence and coronal plasmas, we derive a dispersion relation whose imaginary part describes the evolution of the instability. Results are obtained in the limit of weak dissipation.}
{Using simple analytical methods, we show that dissipative instabilities appear for flow speeds that are lower than the Kelvin-Helmholtz instability threshold. While viscosity tends to destabilise the plasma, the effect of partial ionisation (through the Cowling resistivity) will act towards stabilising the interface. For ionisation degrees closer to a neutral gas the interface will be unstable for larger values of equilibrium flow. The same principle is assumed when studying the appearance of instability at the interface between prominences and dark plumes. The unstable mode appearing in this case has a very small growth rate and dissipative instability cannot explain the appearance of flows in plumes.}
{The present study improves our understanding of the complexity of dynamical processes at the interface of solar prominences and solar corona, and the role partial ionisation can have on the stability of the plasma.  Our results clearly show that the problem of partial ionisation introduces new aspects of plasma stability with consequences on the evolution of solar prominences. }


\maketitle

\section{Introduction}

Solar prominences are among the most enigmatic structures in the solar atmosphere whose study is made difficult by their complex evolution and the multitude of important effects appearing in them. Prominences are believed to be of chromospheric origin and some of them show a long-term stability.  When formed, prominences maintain their high density and low temperature despite being surrounded by the million degree solar corona. Their stability and thermal shielding is provided by the magnetic field. Their importance resides in the recognition that almost 80\% of the observed coronal mass ejections (CMEs) - believed to drive the space weather- have a cold chromospheric core believed to originate from a prominence, which is why the study of the generation and evolution of prominences is necessary.

The difficulty in studying prominences arises from the complex dynamics occurring in these magnetic features, but also because of their intrinsic structure and properties. Early observations showed that prominences are made up of fine structures that are composed of many horizontal, thin dark threads (filaments) (e.g. de Jager 1959, Kuperus and Tandberg-Hanssen 1967) with average width of 200 km and lengths of 3,500-28,000 km  (see e.g. the review by Lin 2010 and references therein). These threads are tracers of the magnetic field. 

Recent H$_{\alpha}$ and UV/EUV observations showed that solar prominences are also very dynamic with observed bulk flows in the range of 2-35 km s$^{-1}$ (e.g. Berger et al. 2008).  In active region prominences, flow velocities seem to be higher than in quiescent prominences, even reaching 200 km s$^{-1}$, and some of these high-speed flows are probably related
to the prominence formation itself. The
range of observed velocities of filament flows is between 5 and 20 km s$^{-1}$. A particular feature in
these observations is the presence of counter-streaming flows, i.e. oppositely directed flows (Zirker
et al., 1998; Lin et al., 2003). Because of the physical conditions of the filament plasma, all these
flows seem to be field-aligned. For a detailed review of the observed flows in solar prominences see Labrosse et al. (2010) and Mackay et al. (2010).

Significant advancement in the study of prominences was made when high-resolution observations  of waves, oscillations, and flows became available. Scientists were able to connect theoretical models with observations  through seismological techniques in order to derive quantities and processes (structure of the magnetic field, transport mechanisms acting in prominences, internal structure, etc.) that cannot be measured directly or indirectly (for a detailed discussion of seismological techniques and results see the review by Arregui et al. 2012). There is also some evidence that velocity oscillations are more easily detected at the edges of prominences or where the material seems fainter, while they are sometimes harder to detect at the prominence main body (Tsubaki and Takeuchi, 1986; Tsubaki et al., 1988; Suematsu et al., 1990; Thompson and Schmieder, 1991; Terradas et al., 2002). 

One of the fundamental properties of solar prominences is that because of their relatively low temperature, the plasma is not fully ionised and  its description therefore needs  special attention, especially when compared to the fully ionised coronal plasma that surrounds prominences. The ionisation degree of prominences is not well known, but there is plentiful evidence that this cannot be neglected when one studies the dynamics and stability of these structures (Patsourakos and Vial 2002, Gilbert et al. 2007, Labrosse et a. 2010, Zaqarashvili et al. 2011, Khomenko and Collados 2012, etc.). 

The  problem of prominence stability is paramount for other effects such as CME eruption due to the connection between these two solar atmospheric structures. In a recent series of papers Ryuota et al. (2010), Berger et al. (2010), Terradas et al. (2012) highlighted a number physical processes taking place in solar prominences that can be connected to instabilities, such as Rayleigh-Taylor instabilities  (RTI) and Kelvin-Helmholtz instabilities (KHI)  under the effect of plasma flows. The effect of partial ionisation on the stability of prominences was investigated earlier by e.g. Diaz et al. (2012), who analysed the appearance of RTI in partially ionised prominence plasma.  These authors found that the linear growth rate is lowered by both the compressibility of the gas and ion-neutral collisions, even though the appearance threshold of this instability is not altered. They also found that the ion-neutral collisions have a strong impact on the RTI growth rate, which can be decreased by an order of magnitude compared to the case corresponding to the collisionless limit. They conclude that their results could explain the existence of prominence fine structure with lifetimes of the order of 30 minutes, a duration that classical theories cannot explain.

In the same year, Soler et al. (2012) investigated the KHI of compressional and partially ionised prominence plasma. They considered the stability of an interface separating two partially ionised plasmas in the presence of a shear flow. In the incompressible limit the KHI was present for any value of the flow, regardless the degree of ionisation. When extended to a compressible limit, the instability threshold was very much sensitive to the collision frequency and density contrast between the two layers of their model. In particular the density contrast is an important parameter in their model. In classical theories the flow speed at which the KHI is set is always super-Alfv\'enic; however, the results of these authors show that for a high density contrast the threshold can be even sub-Alfv\'enic thanks to the ion-neutral coupling. 

In addition to these instabilities there is another, rather unexpected instability that can arise at the interface between two media called {\it dissipative instability} and it is strongly connected to the phenomenon of {\it negative energy waves}. This instability always  occurs for flows lower than the KHI value. Under normal conditions the interface between two media allows the propagation of two modes travelling in opposite directions. For flow speeds larger than a critical value, the propagation direction of the two waves becomes identical, and the wave whose phase speed is smaller becomes a negative energy wave (Ryutova 1988). The dissipative mechanisms acting in the two regions can amplify this negative energy mode leading to {\it dissipative instability}, and the growth rate of this instability is proportional to the combination of dissipative coefficients. Under solar conditions the problem of negative energy waves has been studied by many authors (e.g. Ruderman and Goossens 1995, Ruderman et al. 1996, Joarder et al. 1997, Terra-Homem et al. 2003, etc). In the present study we  consider this problem, but  the two regions separated by the interface are the viscous corona and the partially ionised prominence plasma.

The concept of negative energy waves is based on the energy equation
\[
\frac{dE}{dt}=-D,
\]
where $E$ is the linear part of the energy and $D$ is the dissipative function. The two functions appearing in the above relation depend on the choice of the frame of reference. If we choose a coordinate system where $D>0$, then the variation of the energy with time is negative, meaning that the energy of the system decays as a result of dissipation. In this case $E>0$ for positive energy waves, and dissipation leads to the damping of the wave, i.e. to a decay in its amplitude. However, if $E<0$ the wave is called a negative energy wave and dissipation leads to an amplification of the wave amplitude resulting in an instability. 

The paper is organised as follows. In Sect. 2 we introduce the basic assumptions regarding the nature of partial ionisation and discuss the dissipative mechanisms applied in our study. Here we also introduce the mathematical framework we employ when studying the stability of the plasma. In section 3 we derive the dispersion relation of incompressible waves propagating at the interface between the viscous coronal plasma and partially ionised prominence and establish the stability thresholds of this model. Finally, our results are summarised in section 4.

\section{Governing equations and basic assumptions}

\subsection{Equilibrium}

We assume two semi-infinite layers of collisional and incompressible plasma separated by an interface modelling the interface between the solar prominence and solar corona.

The interface between the corona (labelled with index ``1'') and the solar prominence (labelled with index ``2'') is situated at $z=0$ in a two-dimensional ($x-z$) cartesian reference system. The homogeneous magnetic field in both regions is along the $x$-axis with $B_{01}\neq B_{02}$. The unperturbed state is characterised by a magnetohydrodynamics (MHD) tangential discontinuity at
$z = 0$, and all equilibrium quantities are constant at both sides of the discontinuity. We assume that there is an equilibrium flow in the positive $x$ direction in the prominence (for the $z>0$ region), while in the corona (corresponding to $z<0$) the equilibrium is static. The above equilibrium describes the interface between a prominence and the surrounding quiet corona. Although these two solar regions have been neighbours  for a very long time (in the case of quiescent prominences their stability is shown to be of the order of several months) they present a very different set of physical parameters describing them. Solar prominences are cool and dense plasma material, thought to be of chromospheric origin that are surrounded by the very hot and very tenuous solar corona. Accordingly, it is customary to consider that the density of the prominence is two orders of magnitude larger than the density of the corona and the temperature two orders of magnitude lower than the coronal temperature. Gravity is neglected and so the RTI is not present in our problem. 

\subsection{Basic assumptions}

The determination of transport mechanisms acting in solar plasmas is a very difficult task. After all, the dominant dissipative mechanism depends not only on the location where the dynamics occurs, but also on the nature of the physical mechanisms that needs describing. Under prominence conditions Ballai (2003) and Carbonell et al. (2004) showed that none of the classical dissipative processes (assuming a fully ionised plasma) are able to describe realistic damping of observed waves in prominences, except thermal conduction. Recent studies (e.g. Khodachenko et al. 2004, Arber et al. 2007, Forteza et al. 2007, 2008) also showed that the dominant transport mechanism in solar prominences is probably due to the partially ionised  character of the plasma. Soler et al. (2009b) found that resonant absorption is dominant over ion-neutral effects in the damping of the kink mode in prominence threads. In the present study the appearance of resonant absorption is prevented by assuming a sharp transition between the prominence and corona. 

In partially ionised plasmas the classical Coulomb resistivity is several orders of magnitude smaller than the Cowling resistivity, and the viscosity of the plasma is provided by the friction between various particles making up the plasma (neutrals, ions, protons). The second consideration also implies that the dynamics in solar prominences has to be described in a multi-fluid plasma. However, if the resistivity of the plasma is dominant (as is assumed here) the plasma is described within the framework of single-fluid MHD. In the present paper we will assume that these restrictive conditions are satisfied, i.e. we are going to use a single-fluid description.

In  partially ionised prominence plasmas the Coulomb resistivity is many orders of magnitude smaller than the Cowling anisotropic resistivity (see e.g. Cowling 1957, Khodachenko et al. 2004). Indeed, their difference is given by 
\begin{equation}
\eta_C=\eta+\frac{\xi_nB_0^2}{\mu_0\alpha_n}
\label{eq:etac1}
,\end{equation}
where $\eta_C$ is the Cowling resistivity, $\eta$ is the classical Coulomb resistivity, $\mu_0$ is the permeability of free space, $\xi_n$ is fraction of neutrals, and the frictional coefficient $\alpha_n$ in the case of a plasma assumed to be composed entirely of H is given by
\[
\alpha_n=2\xi_n(1-\xi_n)\frac{\rho^2}{m_p}\sqrt{\frac{k_BT}{\pi m_p}}\Sigma_{in},
\]
where $m_p$ is the proton (ion) mass, $\rho$ is density, $k_B$ is the Boltzmann constant,  and $\Sigma_{in}\approx 5\times 10^{-15}$ cm$^{-2}$ is the ion-neutral collisional cross-section. The number densities of electrons and ions are assumed to be approximately equal. The quantity $\xi_n$  plays an important role in our discussion as it contains information about the ionisation degree of the plasma. By definition this quantity reflects the number of neutrals in the gas mixture, i.e.
\begin{equation}
\xi_n=\frac{\rho_n}{\rho}\approx \frac{n_n}{n_i+n_n}.
\label{eq:xin}
\end{equation}
The degree of ionisation can be characterised by the ionisation
fraction (defined as the mean atomic weight, i.e. the average mass per particle in units of proton mass) given as
\begin{equation}
{ \mu}=\frac{1}{2-\xi_n}
\label{eq:mu}
.\end{equation}
According to this definition, a fully ionised gas corresponds
to ${\mu}=0.5$, while a neutral gas is described by ${\mu}=1$. 

Our aim here is to study the appearance and evolution of instabilities at the interface of two media, therefore we neglect the effects of particle
ionisation and recombination in the solar prominence. Here we assume a strong
thermal coupling between the species, which means electrons,
ions, and neutrals have the same temperature (i.e.
$T_e=T_i=T_n=T$). Therefore, the three-component  gas
can be considered as a single fluid. The concept of a three-component gas mixture will introduce new types of transport
mechanisms whose importance in the context of solar
prominences was discussed in detail in the pioneering work
of Forteza et al. (2007). Since we are going to limit ourselves
to linear dissipation, we will neglect effects connected to
the inertia of different particles, but also the transversal
drift of charged particles due to the Hall term and consider
that thermodynamic quantities (pressure, temperature) are
relatively smooth functions of the spatial coordinates, i.e.
the relative densities of neutrals and ions are constants.
Therefore, when describing the dynamics in solar prominences we will restrict our model to transport mechanisms
that arise in the induction equation.

Temperatures in the solar corona can reach millions of degrees K, therefore the plasma can be considered to be completely ionised. In this important solar region the product $\omega_{ci}\tau_i\gg 1$ (where $\omega_{ci}$ is the ion cyclotron frequency and $\tau_i$ is the ion mean collisional time), therefore ions can gyrate many times around magnetic field lines between collisions. Under typical coronal conditions this product is of the order of $10^5$. Provided the characteristic scales are larger than the mean free path of ions, viscosity in the solar corona is mainly due to ions and the viscosity gyrating around the magnetic field is given by the Braginskii stress tensor (Braginskii 1965) whose linearised expression takes the form of a sum of five terms each with different physical meaning (see e.g.  Erd\'elyi \& Goossens 2004; Ruderman et al. 2000, Mocanu et al. 2008). Under coronal conditions the first term, called parallel or compressional viscosity, is dominant (by several orders of magnitude) and controls the variation along magnetic field lines of the velocity component parallel to field lines. The parallel viscosity is due to the collision-induced random-walk diffusion of particles and is given by
\[
\eta_0=\frac{\rho_0T_0k_B\tau_i}{m_p},
\]
where $\rho_0$ and $T_0$ are the density and temperature of the medium. In practice it is more convenient to work with the kinematic coefficient of viscosity defined as $\nu=\eta_0/\rho_0$.
As determined by Ruderman et al. (1996), a property of the highly anisotropic viscosity is that it allows a jump in the velocity across a magnetic surface, since a strong magnetic field causes ions to rotate around the magnetic field lines, thus preventing the diffusion of particles across the field lines. This also implies that there is no momentum transport across the magnetic surfaces, and different layers  of plasma can move  with respect to each other without friction.  

Observations show that quiescent prominences are made of chromospheric material and they live in a relatively stable position for a long time. High-resolution observations very often  show that the edges of prominences are not still; small- and large-scale features appear and disappear rather frequently. We will try to explain these modifications in the interface between the two media by instabilities that develop owing to the amplification of waves propagating along the interface. 

\subsection{Governing linearised equations}

We will consider a two-layer system, where an interface separates the solar prominence and solar corona. The equations describing the dynamics of the plasma are the incompressible dissipative and linear MHD equations. In both regions the equations
\begin{equation}
\nabla \cdot {\bf v}=0,\quad \nabla \cdot {\bf b}=0,
\label{eq:1.1}
\end{equation}
are valid. In the solar prominence we assume a field-aligned equilibrium flow (${\bf v_0}$). As a result, the momentum equation becomes
\begin{equation}
\rho_2\frac{\partial {\bf v_2}}{\partial t}+v_0\frac{\partial {\bf v_2}}{\partial x}=-\nabla P_2+\frac{B_{02}}{\mu_0}\frac{\partial {\bf b_2}}{\partial x}.
\label{eq:1.3}
\end{equation}
In the solar corona the equilibrium is static, but the momentum equation is supplemented by the viscous force, i.e.
\begin{equation}
\rho_1\frac{\partial {\bf v_1}}{\partial t}=-\nabla P_1+\frac{B_{02}}{\mu_0}\frac{\partial {\bf b_1}}{\partial x}+{\bf \cal V}.
\label{eq:1.3.1}
\end{equation}
In the solar prominence the dominant dissipative effect is the Cowling resistivity, therefore the induction equation becomes
\begin{equation}
\frac{\partial {\bf b_2}}{\partial t}+v_0\frac{\partial {\bf b_2}}{\partial x}=B_{02}\frac{\partial {\bf v_2}}{\partial x}+{{\bf \cal R}}.
\label{eq:1.4}
\end{equation}
In the corona, the flow and dissipative effects do not modify the induction equation, so we can write
\begin{equation}
\frac{\partial {\bf b_1}}{\partial t}=B_{01}\frac{\partial {\bf v_1}}{\partial x}.
\label{eq:1.4.1}
\end{equation}
In the above equations ${\bf v_i}$ and ${\bf b_i}$ ($i=1, 2$) are the perturbations of velocity and magnetic field, $P_i$ are the total pressures (the sum of kinetic and magnetic pressures), and the dissipative terms in Eqs. (\ref{eq:1.3.1}) and (\ref{eq:1.4}) are given by (see e.g. Ruderman et al. 1996)
\[
{{\bf \cal V}}=\nu\left[{\bf \tilde b}({\bf \tilde b}\cdot \nabla)-\frac13\nabla\right]\left[{\bf \tilde b}\cdot \nabla ({\bf \tilde b}\cdot {\bf v_1})\right],
\]
\begin{equation}
{{\bf\cal R}}=\eta\nabla^2{\bf b}_2+\frac{(\eta_C-\eta)}{|{\bf B}_0|^2}\nabla\times\left\{\left[\left(\nabla\times {\bf b_2}\right)\times {\bf B}_0\right]\times {\bf B}_0\right\},
\label{eq:1.5}
\end{equation}
where ${\bf \tilde b}$ is the unit vector in the direction of the magnetic field, i.e. ${\bf \tilde b}={\bf B}_0/B_0$. 

Because of the orientation of the equilibrium magnetic field the interface between the corona and solar prominence can be considered a tangential discontinuity. We write the equation of the perturbed interface as $z=\zeta(x,t)$. We assume that at $|x|\rightarrow \infty$ and $|z|\rightarrow \infty$ all perturbations vanish. At the interface surface waves will be able to propagate, as suggested in an earlier investigation by Roberts (1981). According to his results, in the absence of any equilibrium flow, incompressible Alfv\'enic waves can propagate with a phase speed that lies between the Alfv\'en speeds in the two regions, which is given by
\begin{equation}
\frac{\omega}{k_{x}}=\pm\left(\frac{\rho_1v_{A1}^2+\rho_2v_{A2}^2}{\rho_1+\rho_2}\right)^{1/2}=\pm\left(\frac{v_{A1}^2+dv_{A2}^2}{1+d}\right)^{1/2},
\label{eq:1.6}
\end{equation}
where $d=\rho_2/\rho_1$ is the density contrast parameter, $k_{x}$ is the parallel component of the wavevector to the interface, and $v_{A1}=B_{01}/\sqrt{\mu_0\rho_1}$ and $v_{A2}=B_{02}/\sqrt{\mu_0\rho_2}$ are the Alfv\'en speeds in the two regions. This dispersion relation describes the propagation of the two waves along the interface in opposite directions.  

For a stable interface at $z=0$ we have to impose the linearised kinetic boundary condition and the condition of the continuity of normal component of stresses. If ${\bf v}_i=(v_{xi},\;0,\;v_{zi})$, then these conditions read
\begin{equation}
v_{z1}=\frac{\partial \zeta}{\partial t}, \quad v_{z2}=\frac{\partial \zeta}{\partial t}+v_0\frac{\partial \zeta}{\partial x},
\label{eq:1.7}
\end{equation}
and
\begin{equation}
P_1+\rho_1\nu{\bf \tilde b}\cdot\nabla ({\bf \tilde b}\cdot {\bf v})\equiv P_1-\rho_1\nu\frac{\partial v_{z1}}{\partial z}=P_2.
\label{eq:1.8}
\end{equation}
We note here that at the tangential discontinuity used in the present paper there is no mass transfer between the two media, meaning that the state of the plasma in each region is not disturbed by the presence of the other medium. The system of equations (\ref{eq:1.1})--(\ref{eq:1.5}) together with the boundary conditions at the interface will form the starting equations for our discussion on dissipative instability generated at the interface between the two media.

\section{Dispersion relation of surface waves at the discontinuity}

Since we are going to deal with an eigenvalue problem we will perform a normal mode analysis and take all perturbations proportional to $\exp[i(k_xx-\omega t)]$, where $\omega$ is a complex frequency that can be written as $\omega=Re(\omega)+iIm(\omega)$. This particular form of perturbations reduces the boundary conditions to
\begin{equation}
v_{z1}=-i\omega\zeta, \quad v_{z2}=-i\Omega\zeta,
\label{eq:1.8.1}
\end{equation}
where $\Omega=\omega-k_xv_0$ is the Doppler-shifted frequency of waves in the solar prominence. 

When computing the components of the resistive terms given by Eq. (\ref{eq:1.5}) together with the solenoidal condition (\ref{eq:1.1}) we can obtain that all dissipative terms containing the classical Coulomb resistivity cancel, therefore the dissipation in the partially ionised prominence is described by the Cowling resistivity alone. 

We introduce the viscous and resistive Reynolds numbers as
\begin{equation}
R_v=\frac{v_{A1}}{k_x\nu}, \quad R_{r}=\frac{v_{A2}}{k_x\eta_C}.
\label{eq:2.1}
\end{equation}
Under coronal and prominence conditions both Reynolds numbers are very large and therefore waves will propagate with little damping over a period, meaning that in our subsequent calculations we will consider that $|Re(\omega)|\gg|Im(\omega)|$. The very large Reynolds numbers also allow us to consider dissipative terms  much smaller than other terms belonging to ideal MHD, meaning that in our calculations all terms containing $\nu^2$ or $\eta_C^2$ are neglected. The interaction between flows and waves propagating at the interface between the two media is ensured by dissipation that could play the role of energy sink. Later we will see that, contrary to our expectations, dissipation does not always act towards decreasing the wave amplitude; for specific values of flows or ionisation degree, the interplay between flows, dissipation, and waves could lead to an increase of the waves' amplitude, i.e. unstable behaviour.

In  region 1 the viscous MHD equations can be reduced to a system of coupled equations for the normal component of the velocity vector $v_{z1}$ and total pressure $P_1$ of the form
\begin{equation}
\frac{dv_{z1}}{dz}=-\frac{ik_x^2\omega}{\rho_1({\cal D}_{A1}+2i\nu k_x^2\omega)}P_1,
\label{eq:2.2}
\end{equation}
\begin{equation}
\left(1-\frac{i\nu\omega}{{\cal D}_{A1}}\frac{d^2}{dz^2}\right)v_{z1}=-\frac{i\omega}{\rho_1{\cal D}_{A1}}\frac{dP_1}{dz},
\label{eq:2.3}
\end{equation}
where ${\cal D}_{A1}=\omega^2-k_x^2v_{A1}^2$. Taking into account that $R_v\gg 1$, we can eliminate the total pressure from these two equations to arrive at a single relation for $v_{z1}$, i.e.
\begin{equation}
\frac{d^2v_{z1}}{dz^2}-k_x^2\left(1-\frac{3i\nu k_x^2\omega}{{\cal D}_{A1}}\right)v_{z1}=0.
\label{eq:2.4}
\end{equation}
It is easy to see that $v_{z1}$ will vanish as $z\rightarrow -\infty$. In order to use the boundary conditions (\ref{eq:1.7}) and (\ref{eq:1.8}) we will also need to find  the value of the total pressure. In order to calculate its expression we write the equation for the $z$-component of the velocity (\ref{eq:2.4}) as
\begin{equation}
\frac{d^2v_{z1}}{dz^2}-\alpha^2v_{z1}=0,
\label{eq:2.5}
\end{equation}
where 
\[
\alpha=k_x\left(1-\frac{3i\nu k_x^2\omega}{{\cal D}_{A1}}\right)^{1/2}\approx k_x\left(1-\frac{3i\nu k_x^2\omega}{2{\cal D}_{A1}}\right).
\]
Equation (\ref{eq:2.5}) allows us to explicitly find the expression of the $z$-component of the velocity in the solar corona. With the help of Eqs. (\ref{eq:1.7}) and (\ref{eq:2.2}) we can find that the total pressure in region 1 estimated at the interface between the two regions can be written as
\begin{equation}
P_1=\frac{\rho_1D_{A1}}{k_x}\left(1-\frac{i\nu\omega k_x}{2D_{A1}}\right)\zeta.
\label{eq:2.6}
\end{equation}

For the prominence we now have an equilibrium flow in the positive $x$ direction. Considering again the equations that relate the $z$-component of the velocity vector and total pressure we obtain the systems
\begin{equation}
\left({\cal D}_{A2}+\frac{i\eta_Ck_x^4v_{A2}^2}{\Omega}\right)v_{z2}=-\frac{i\Omega}{\rho_2}\frac{dP_2}{dz}
\label{eq:2.10}
\end{equation}
and
\begin{equation}
\left({\cal D}_{A2}+\frac{i\eta_Ck_x^4v_{A2}^2}{\Omega}\right)\frac{dv_{z2}}{dz}=-\frac{i\Omega}{\rho_2}P_2,
\label{eq:2.11}
\end{equation}
where ${\cal D}_{A2}=\Omega^2-k_x^2v_{A2}^2$. Eliminating the total pressure from the above two expressions we obtain an equation for $v_{z2}$ valid in the solar prominence of the form
\begin{equation}
\frac{d^2v_{z2}}{dz^2}-k_x^2v_{z2}=0.
\label{eq:2.12}
\end{equation}

Using Eqs. (\ref{eq:1.7}) and (\ref{eq:2.11}) we can write that the total pressure at the prominence evaluated at the interface behaves as
\[
P_2=
\]
\begin{equation}
-\frac{\rho_2(D_{A2}+i\eta_C\Omega k_x^2)\Omega\zeta}{\Omega+i\eta_Ck_x^2}\approx-\frac{\rho_2\zeta}{\Omega}(D_{A2}\Omega+ik_x^4v_{A2}^2 \eta_C).
\label{eq:2.14}
\end{equation}
The expressions of the two total pressures in the two regions are inserted in Eq. (\ref{eq:1.8}), which   leads to the dispersion relation of the form
\begin{equation}
D(\omega)=D_r+iD_i=0,
\label{eq:2.15}
\end{equation}
where 
\[
D_r=D_{A1}+dD_{A2},
\]
\begin{equation}
D_i=\nu k_x^2\omega+\frac{dk_x^4\eta_Cv_{A2}^2}{\Omega}.
\label{eq:2.16}
\end{equation}
In deriving the dispersion relation (\ref{eq:2.15}) we took into account that a perturbation method is used meaning that terms proportional to $\nu^2$ and $\eta_C^2$ are neglected.

\subsection{Instability conditions}

Since we assumed that the damping of waves propagating along the interface is weak, we can write the frequency of waves as $\omega=Re(\omega)+iIm(\omega)$ with $|Re(\omega)| \gg |Im(\omega)|$. This assumption is in line with our previous statement regarding the high Reynolds numbers of solar plasmas and the working supposition that terms containing squares and products of dissipative coefficients can be neglected.
According to the dependence of perturbations on the variable $t$ assumed earlier, $Im(\omega)>0$ corresponds to an overstability, i.e. to a  situation where the amplitude of waves propagating along the interface grows as $\exp(\omega_it)$. Following the method developed by Cairns (1979) we write the dispersion relation as
\begin{equation}
D_r(\omega,k_x)=-i\nu k_x^2\omega-\frac{idk_x^4\eta_Cv_{A2}^2}{\Omega}.
\label{eq:2.17}
\end{equation}
The solution of the equation $D_r=0$ will result in the real part of the frequency $\omega$. In ideal MHD the interface between the two regions is always stable; however, the introduction of dissipation may lead to instability. The dispersion relation for the ideal case can be easily solved and leads to the frequency 
\begin{equation}
Re(\omega)^{\pm}=\frac{k_xv_0d}{1+d}\pm \frac{k_x}{1+d}\Big[d(v_{KH}^2-v_0^2)\Big]^{1/2}.
\label{eq:2.18}
\end{equation}
Equation (\ref{eq:2.18}) describes two waves propagating along the interface in opposite directions. The quantity $v_{KH}$ is the Kelvin-Helmholtz (KH) threshold velocity given by
\begin{equation}
v_{KH}^2=\frac{1+d}{d}(v_{A1}^2+dv_{A2}^2),
\label{eq:2.19}
\end{equation} 
and it plays a very important role in the discussion of stability of waves propagating in a flowing plasma. It is obvious from Eq. (\ref{eq:2.18}) that the plasma becomes KH unstable for flows that satisfy the condition $v_0^2>v_{KH}^2$. We estimate the value of $v_{KH}$ for the system under  investigation. If we consider typical coronal and prominence values for density and Alfv\'en speeds ($d=100$, $v_{A1}=315$ km s$^{-1}$, $v_{A2}=28$ km s$^{-1}$, all taken from Joarder and Roberts 1992) we obtain $v_{KH}=423$ km s$^{-1}$. It is obvious that observations in the solar prominences do not show equilibrium flows that are larger than $v_{KH}$; in reality, these speeds are more likely to be of the order of a tenth of $v_{KH}$ or smaller. This means that under prominence conditions the plasma at the interface between the solar corona and solar prominences is always KH stable. In the absence of any flow, the two solutions of Eq. (\ref{eq:2.18}) describe two modes propagating in the opposite direction with equal speeds $\left|v_{KH}\sqrt{d}/(1+d)\right|$. In the presence of a flow (for our problem the flow is present in the prominence while the coronal plasma is at rest), waves are drifted by the flow. Since the flow direction points in the positive direction, the flow affects the two waves in a different way and the symmetry of the two modes is lost. It can be easily shown that the difference between the phase speeds of the two waves is $2v_0d/(1+d)$. For flow speeds larger than $v_{KH}/\sqrt(1+d)$ the direction of the wave propagating in the negative direction is inverted and the backward mode becomes the forward mode. The two modes can amplify each other leading to instability. If we plot the two frequencies obtained in Eq. (\ref{eq:2.18}) with respect to an increasing flow speed we  obtain that the KHI occurs when the oscillation frequencies of the forward and backward propagating surface modes merge for increasing flow velocity. The
merging point then indicates the threshold of KHI for the single interface. In the present analysis we  consider flows that are less than the KH threshold. 

Since we assume that the damping is weak, the imaginary part of the frequency can be given by
\begin{equation}
Im(\omega)\approx -\frac{k_x^2}{\partial D_r/\partial Re(\omega)}\left( \nu \omega_r+\frac{dk_x^2\eta_Cv_{A2}^2}{\Omega_r}\right),
\label{eq:2.19.1}
\end{equation}
where $\Omega_r=Re(\omega)-k_xv_0$.

Using Eqs. (\ref{eq:2.17}) and (\ref{eq:2.18}), it is straightforward to show that the imaginary part of the frequencies are
\begin{equation}
Im(\omega)^{+}=-\frac{\nu k_x^2}{2(1+d)}\left(\frac{v_0d}{\Gamma}+ 1\right)+ \frac{d(d+1) k_x^2v_{A2}^2\eta_C}{2(v_0\Gamma- d\Gamma^2)},
\label{eq:2.20}
\end{equation}
and 
\begin{equation}
Im(\omega)^{-}=\frac{\nu k_x^2}{2(1+d)}\left(\frac{v_0d}{\Gamma}-1\right)- \frac{d(d+1) k_x^2v_{A2}^2\eta_C}{2(v_0\Gamma+d\Gamma^2)},
\label{eq:2.20.1}
\end{equation}
where $\Gamma=\sqrt{d(v_{KH}^2-v_0^2)}$. With the help of Eqs. (\ref{eq:etac1})-(\ref{eq:mu}) we can write the Cowling resistivity as
\begin{equation}
\eta_C=\frac{v_{A2}^2m_n(2{\mu}-1)}{2\rho_2(1-{\mu})\Sigma_{in}}\left(\frac{\pi m_p}{k_BT_{2}}\right)^{1/2}.
\label{eq:2.21}
\end{equation}
As a result, the two values for the imaginary part of the frequency become
\[
Im(\omega)^{+}=-\frac{\nu k_x^2}{2(1+d)}\left(\frac{v_0d}{\Gamma}+1\right)+
\]
\begin{equation}
\frac{d(d+1)k_x^2m_n(2{\mu}-1)v_{A2}^4}{4(v_0\Gamma-\Gamma^2)(1-{\mu})\rho_2\Sigma_{in}}\left(\frac{\pi m_p}{k_BT}\right)^{1/2}
\label{eq:2.22}
\end{equation}
and
\[
Im(\omega)^{-}=\frac{\nu k_x^2}{2(1+d)}\left(\frac{v_0d}{\Gamma}-1\right)-
\]
\begin{equation}
\frac{d(d+1)k_x^2m_n(2{\mu}-1)v_{A2}^4}{4(v_0\Gamma+\Gamma^2)(1-{\mu})\rho_2\Sigma_{in}}\left(\frac{\pi m_p}{k_BT}\right)^{1/2}.
\label{eq:2.23}
\end{equation}

We now discuss the sign of these frequencies for a range of flow speeds changing in the interval 10-60 km s$^{-1}$ and for an ionisation degree varied between the cases corresponding to full ionisation (${\mu}=0.5$) and neutral plasma (${\mu}=1$).
A simple graphical analysis clearly shows that for the spectrum of flows considered here and for any ionisation degree the imaginary part of the surface waves that propagates in the positive direction (i.e. in the direction of the flow) is negative leading to a classical physical damping. In contrast the imaginary part of the wave propagating backward (in the negative direction) can become positive for flow speeds larger than 48 km s$^{-1}$ (see Fig. 1). A positive imaginary part of the frequency is connected to an instability.
\begin{figure}
\centering
\includegraphics[width=\columnwidth]{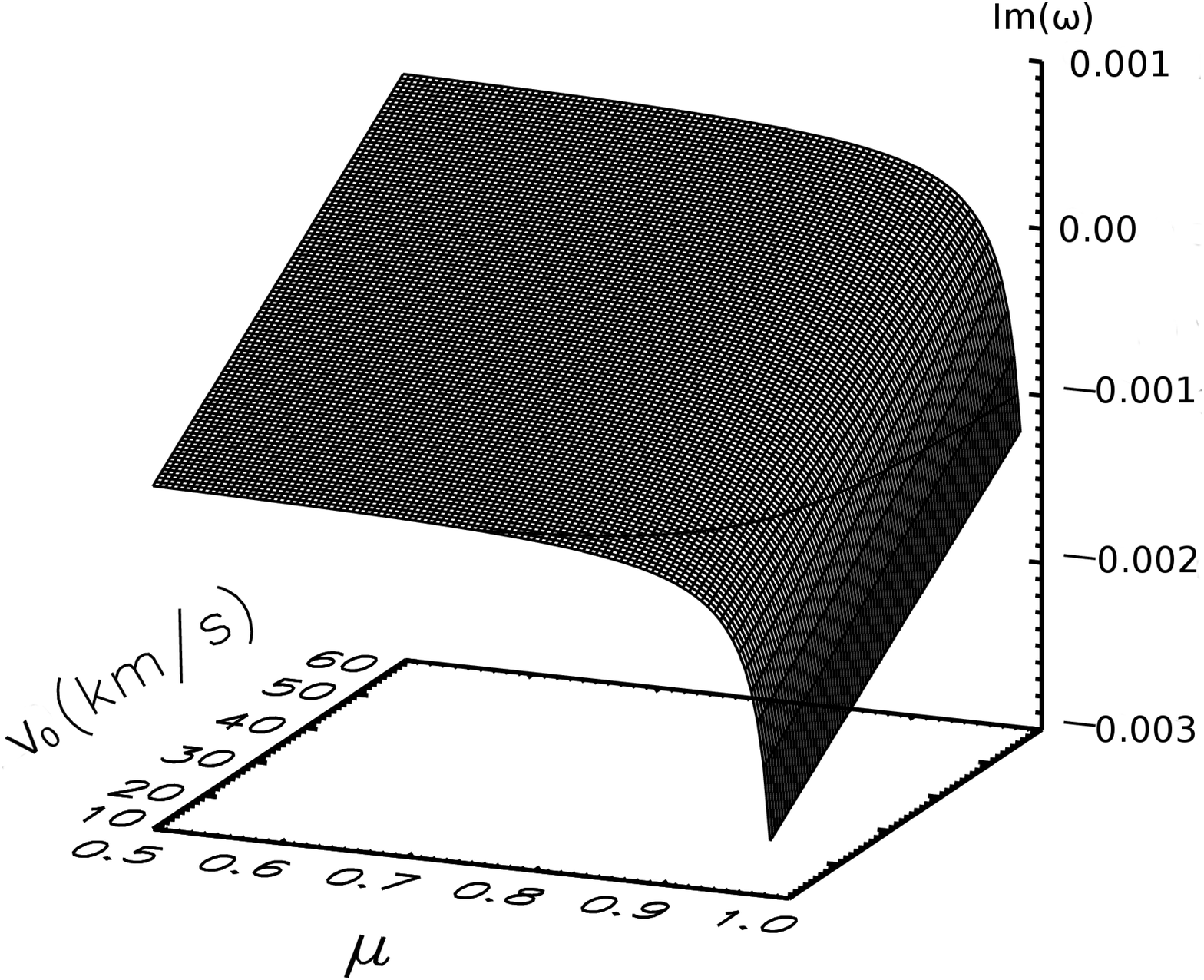}
\caption{Variation of the imaginary part of the frequency for the backward propagating wave with the flow speed and the ionisation fraction. The flow changes in the interval 10-60 km s$^{-1}$ and the ionisation fraction varies between 0.5 (fully ionised plasma) and 1 (neutral gas). The horizontal curve is drawn at the $Im(\omega)=0$ and helps to visualise the transition of $Im(\omega)^{-}$ from the positive to the negative domain.}
\label{fig3}
\end{figure}
A contour plot of the imaginary part of the frequency for the backward propagating wave is shown in Fig. 2, where the role of the partial ionisation and plasma flows becomes evident. The region above the $\Im(\omega)=0$ curve corresponds to the region where the wave is unstable, while in the region beneath the curve the wave is stable and damped. It is clear that the flow will destabilise the interface; for a given value of ionisation fraction there is a flow value at which the interface becomes unstable (a similar conclusion can be drawn from earlier studies by e.g. Ruderman and Goossens 1995). The variation of the zero-level with respect to the ionisation fraction shows that as the plasma  becomes more dominated by neutrals, the plasma interface becomes more stable, so that for an ionisation degree of 0.93 the interface becomes stable and waves will damp owing to dissipation. Figure \ref{fig3} also shows that the presence of neutrals stabilises the plasma as the instability sets in for higher values of flows (here with a density ratio of 100, $k_x=5\times 10^{-6}$ m$^{-1}$, $\nu=10^{10}$ m$^2$ s$^{-1}$, $T=10^4$ K, $\rho_2=5\times 10^{-11}$ Kg m$^{-3}$). 
\begin{figure}
\centering
\includegraphics[width=\columnwidth]{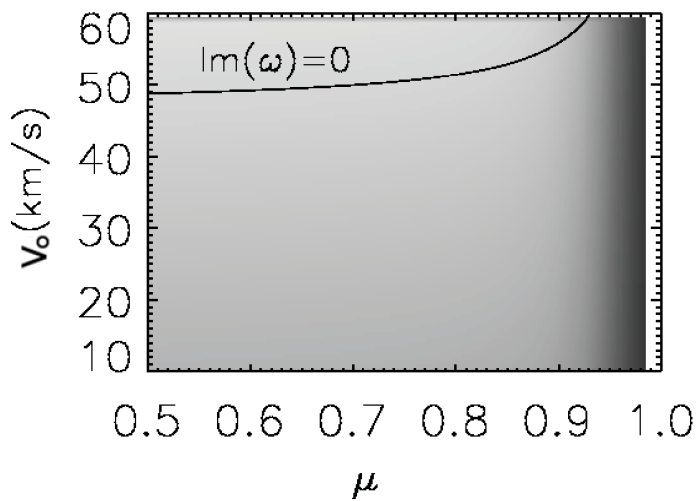}
\caption{ Contour plot of the variation of the imaginary part of the frequency for {\it backward} propagating waves. The region below the zero level curve corresponds to a stable regime and waves will have a classical damping, while the interface described by the quantities in the region above the curve is unstable.} 
\label{fig4}
\end{figure}
It is instructive to identify the role of each dissipative process in the appearance of instability. While the partial ionisation in the solar prominence has the effect of stabilisation of the interface, the viscosity in the solar corona will destabilise the discontinuity and the value of the flow at which waves become unstable has little variation with the ionisation fraction and significant dependence can be observed for larger values of ${\mu}$. We note that the unstable behaviour of the backward wave is also connected to the very high density contrast between the solar prominence and corona. For a density contrast of one order of magnitude the unstable backward wave becomes stable and the imaginary part of the dispersion relation describes classical physical damping. 

Finally, we explore the connectivity between the dissipative instability discussed earlier and negative energy waves. As specified in the Introduction, the term of negative
energy wave refers to the situation when the wave energy decreases with the increase of the wave
amplitude. The energy of a wave with amplitude $A$ averaged over one wavelength can be given as
\[
E=\frac14\omega\frac{\partial D_r}{\partial \omega}|A|^2,
\]
where $D_r$ is the dispersion relation of the wave. In this case the energy of the wave is the phase-averaged difference between the energy of the system when the wave is present, and its energy when the wave is absent. 
A criterion that can be used to determine the nature of waves is the formula suggested by Cairns (1979) where a wave is considered to have negative energy if the quantity
\begin{equation}
C=Re(\omega)\frac{\partial D_r}{\partial\omega}<0.
\label{eq:2.25}
\end{equation}
The function $D_r$ is undetermined up to a multiplicative constant whose sign has to be determined from the condition that in the absence of any flows in the system $C>0$. Comparing this with Eq. (\ref{eq:2.17}), it is obvious that the condition for the appearance of dissipative instability is identical with the condition of negative energy wave generation because the expression
\[  
 \nu Re(\omega)^2+\frac{dk_x^2\eta_Cv_{A2}^2Re(\omega)}{\Omega_r}
 \]
is always positive.

Another possibility for exploiting the effect of partial ionisation on the stability at a magnetic interface is to model the interface between two partially ionised plasmas of prominences and dark plumes. Observations by Berger et al. (2010) revealed the existence of dark plumes within the prominence showing turbulent upflows in prominences of the order of 15-30 km s$^{-1}$. These upflows are believed to generate instabilities. In the Ca II H-line plumes are
seen dark in contrast to the prominence material, which suggests
that the plasma in the plumes is hotter and probably less
dense than the prominence material. The width of the plumes
ranges between 0.5 Mm and 6 Mm, and their maximum heights
are between 11Mm and 17 Mm. The typical plume lifetime is between 400 s and 890 s. 

Considering the same prominence/plume parameters as in Soler et al. (2012), we obtain that the interface between these two partially ionised media becomes unstable for almost all values of flows (below the KH threshold) for an ionisation degree of the prominence larger than the ionisation degree of the plume, but the growth rate of this instability is very low, meaning that the dissipative instability (at least in this simplified framework) cannot explain the generation of upflows in plumes by instability.

\section{Conclusions}

In the present study we explored the stability of a tangential discontinuity by modelling the interface between the viscous and fully ionised coronal plasma and the partially ionised solar prominence in which the dominant dissipative effect is the Cowling resistivity. The magnetic fluids were assumed to be incompressible and the prominence equilibrium was considered to be dynamical, with a homogeneous flow parallel to the interface. Assuming a weak damping (confirmed by the very large Reynolds numbers) we obtained the dispersion relation of Alfv\'enic waves propagating along the interface. The presence of dissipative effects on both sides of the interface renders the dispersion relation to be complex with the imaginary part of this quantity describing the decay or the growth of waves. 
Our results show that while the forward propagating wave is always stable, with the amplitude of the wave decaying because of dissipation, for the backward propagating wave there is a threshold of the flow (below the KHI threshold) for which the wave becomes unstable. A careful analysis proves that the partial ionisation has a stabilising effect on the interface for any value of the ionisation fraction and the unstable behaviour can be connected to the viscous nature of the coronal plasma. We also showed that the partial ionisation has little effect on the threshold where waves become unstable. For a plasma where neutrals are abundant, the instability appears for higher values of flows, i.e. neutrals have a stabilising effect. 
The above results were obtained under the strict restriction of incompressible plasma and a sharp tangential discontinuity between the two plasma layers. The same model was used to study the generation of dissipative instability at the interface of two partially ionised plasmas modelling the prominence and dark plumes. The unstable mode obtained in this case shows a very low growth rate, meaning that this type of instability (at least in this simplified model) cannot explain the appearance of turbulent upflows in plumes that can be attributed to instability. The problem of compressibility and a smooth transition between the dynamical solar prominence and static corona in the presence of ion-neutral friction will be addressed in a forthcoming study.

\acknowledgements

The present study was partly carried out while IB was a visiting fellow of UIB. Their hospitality and support is greatly appreciated. IB acknowledges the financial support by the ANCS Romania (STAR, 72/2013). The authors are grateful for the support by the Leverhulme Trust (IN-2014-016)

\end{document}